\documentclass{emulateapj}

\begin{document}

\title{Cool Dust in the Outer Ring of NGC\,1291}

\author{J.~L. Hinz\altaffilmark{1}, C.~W. Engelbracht\altaffilmark{1}, R.\ Skibba\altaffilmark{1}, A.\ Crocker\altaffilmark{2}, J.\ Donovan Meyer\altaffilmark{3}, K.\ Sandstrom\altaffilmark{4}, F.\ Walter\altaffilmark{4}, E.\ Montiel\altaffilmark{1}, B.~D. Johnson\altaffilmark{5}, L.\ Hunt\altaffilmark{6}, G.\ Aniano\altaffilmark{7}, L.\ Armus\altaffilmark{8}, D.\ Calzetti\altaffilmark{2}, D.~A.\ Dale\altaffilmark{9}, B.\ Draine\altaffilmark{7}, M.\ Galametz\altaffilmark{10}, B. Groves\altaffilmark{4}, R.~C.\ Kennicutt\altaffilmark{10, 1}, S.~E.\ Meidt\altaffilmark{4}, E.~J.\ Murphy\altaffilmark{7} and F.\ Tabatabaei\altaffilmark{4}}

\altaffiltext{1}{Steward Observatory, University of Arizona, 933 N.\ Cherry Ave.,  Tucson, AZ  85721, USA; jhinz@as.arizona.edu}

\altaffiltext{2}{Department of Astronomy, University of Massachusetts, Amherst, MA 01003, USA}

\altaffiltext{3}{Physics \& Astronomy Department, Stony Brook University, Stony Brook, NY 11794-3800, USA}

\altaffiltext{4}{Max-Planck-Institut fur Astronomie, Konigstuhl 17, 69117 Heidelberg, Germany}

\altaffiltext{5}{9 Institut d’Astrophysique de Paris, CNRS, Universite Pierre \& Marie Curie, UMR 7095, 98bis bd Arago, 75014 Paris, France}

\altaffiltext{6}{INAF - Osservatorio Astrofisico di Arcetri, Largo E. Fermi 5, 50125 Fireze, Italy}

\altaffiltext{7}{Department of Astrophysical Sciences, Princeton University, Princeton, NJ 08544, USA}

\altaffiltext{8}{Spitzer Science Center, California Institute of Technology, Mc 314-6, Pasadena, CA 91125, USA}

\altaffiltext{9}{Department of Physics \& Astronomy, University of Wyoming, Laramie, WY 82071, USA}

\altaffiltext{10}{Institute of Astronomy, University of Cambridge, Madingley Road, Cambridge CB3 0HA, UK}

\begin{abstract}
 
We examine {\it Herschel Space Observatory} images 
of one nearby prototypical outer ring galaxy, NGC\,1291, and show that the 
ring becomes more prominent at wavelengths longer than 160\,$\micron$.  
The mass of cool
dust in the ring dominates the total dust mass of the galaxy, accounting
for at least 70\% of it.
The temperature of the emitting dust in the ring (T$=19.5\pm0.3$\,K) is cooler 
than that of the inner galaxy (T$=25.7\pm0.7$\,K).  
We discuss several explanations
for the difference in dust temperature, including age and density differences
in the stellar populations of the ring versus the bulge.

\end{abstract}


\section{INTRODUCTION}

Outer or external rings are typically large, low surface brightness features 
of barred and weakly barred galaxies, prominent at optical wavelengths.  They 
are most frequently observed in early-type spirals (S0/a) and, as with gaseous
rings and pseudorings, are most often believed to be associated with
outer Lindblad resonances (Buta \& Combes 1996).  These resonances are assumed
to arise with bars or other pertubations (see simulations by, e.g., 
Bagley et al.\ 2009).  For outer rings in galaxies without bars,
explanations vary from ring creation through tidal 
forcing via interactions with companions or bars that have since 
dissipated, to 
spiral density waves in the disk (Rautiainen \& Salo 2000).  There are also 
outer rings that presumably formed in collisions (e.g., Appleton \& 
Struck-Marcell 1996) or through prograde major mergers of gas-rich disk 
galaxies (Brook et al.\ 2007).

Outer rings can have major axes that are twice the size of the bar component 
(Schwarz 1981) and, therefore, they dominate the outer areas of the
disks of the galaxies that contain them.  The study of outer disks as a path to
understanding galaxy evolution has had recent renewed interest due to the
discovery of very extended disks at ultraviolet (Gil de Paz 2005; Thilker 
et al.\ 2005, 2007; Munoz-Mateos et al.\ 2007), optical (de Jong et al.\ 
2007; McConnachie et al.\ 2009),
infrared (Engelbracht et al.\ 2004; Hinz et al.\ 2004, 2006) and submillimeter
(Planck Collaboration et al.\ 2011) wavelengths.
In particular, implications for the production source of dust and the 
mechanisms for heating
and transporting the dust in galaxy outskirts may reveal much about the growth 
and chemical enrichment of disks.

The proximity of NGC\,1291 and the wealth of available 
ancillary and space-based data
make it an ideal test case for understanding dust emission in these
extended disk structures.
The first large outer ring structure to be discovered was, in fact, 
the one in NGC\,1291 (Perrine 1922).  The galaxy is classified as an 
(R)SB(s)0/a 
(de Vaucouleurs et al.\ 1991) with an inclination of 
$i=35\arcdeg\pm7\arcdeg$ (Prescott et al.\ 2007) and is at a distance 
estimated to be between $7-10.4$\,Mpc (Masters 2005; Kennicutt et al.\ 2008);
for the remainder of the paper we use the 10.4\,Mpc value, following
Kennicutt et al.\ (2011).
In addition to its outer ring, at a radius of $\sim9$\,kpc, at optical 
wavelengths it is characterized by a bright inner lens, a 
primary bar, and a small secondary bar misaligned by $\sim30\arcdeg$ 
(de Vaucouleurs 1975; P{\'e}rez \& Freeman 2006; for more general information
on the classification of rings and lenses in S0's, see, e.g., Michard \& 
Marchal 1993).  Its star formation has been studied via H\,$\alpha$ emission
(Caldwell et al.\ 1991; Crocker et al.\ 1996; Meurer et al.\ 2006).  It is 
part of the {\it Spitzer} Survey of Stellar Structure 
in Galaxies (S$^4$G; Sheth et al.\ 2010) sample and, as such, has been 
re-classified in the Infrared Array Camera (IRAC; Fazio et al.\ 2004) bands as 
(R)SAB(l,ub)0$^+$ (Buta et al.\ 2010).  
NGC\,1291 is easily detected in the ultraviolet, but it is not classified
as exhibiting an extended UV disk because optical emission is found to
be coincident with the UV knots (Thilker et al.\ 2007).  H\,{\sc i}
measurements of the galaxy (van Driel et al.\ 1988) show that the atomic gas is
concentrated in the outer ring with a pronounced central hole.  The 
H\,{\sc i} gas mass is $0.81\times10^9$\,M$_{\odot}$, relatively gas-rich
for an S0/a galaxy (e.g., Li et al.\ 2011).

Using images of the galaxy taken as part of the {\it Spitzer} Infrared
Nearby Galaxies Survey (SINGS; Kennicutt et al.\ 2003) Legacy program, 
Bendo (2006) noted that the nucleus of NGC\,1291 is the dominant
source of 8\,$\micron$ emission, as well as 24 and 70\,$\micron$ warm dust 
emission,
and that the 8 and 24\,$\micron$ emission are well correlated.  However,
at 160\,$\micron$, assumed to be associated with cool (T\,$\sim20$\,K) dust
emission, the outer ring is a stronger source than the central portion 
of the galaxy.  This was confirmed by Balloon-borne Large Aperture 
Submillimeter Telescope ({\it BLAST}; Griffin et 
al.\ 2007) observations of NGC\,1291, presented as part of a paper on 
resolved galaxies that also have {\it Spitzer} images (Wiebe et al.\ 2009).
Both the central core and 
outer ring were detected by {\it BLAST} in a total of four observations of 
a $\sim0.4$\,deg$^2$ area.


In a continuing effort to understand the role dust plays in the
evolution of galaxies, we present {\it Herschel 
Space Observatory} images of NGC\,1291, complementary to the {\it Spitzer}
and {\it BLAST} images described above.  Section 2 describes the observations
and data reduction, and Section 3 the analysis and comparison with previous 
data.  Section 4 contains a discussion of these results, while Section 5 
contains a summary.


\section{OBSERVATIONS AND DATA REDUCTION}
The Multiband Imaging Photometer for {\it Spitzer} (MIPS; Rieke et al.\ 2004)
images used here were observed as part of the SINGS Legacy project
and reduced using 
version 3.06 of the Data Analysis Tool (DAT; Gordon et al.\ 2005).
The DAT performs standard
processing of infrared detector array data (e.g., dark subtraction, flat
fielding) as well as steps specific to the MIPS arrays (droop correction).
The images were calibrated using the most recent values available for
MIPS data at all three wavelengths (Engelbracht et al.\ 2007; Gordon et al.\ 
2007; Stansberry et al.\ 2007).
The MIPS spatial resolutions are 6$\arcsec$, 18$\arcsec$,
and 40$\arcsec$ at 24, 70, and 160\,$\micron$, respectively.

Observations of NGC\,1291 with the {\it Herschel Space Observatory}
were taken as part of the Key Insights on Nearby Galaxies: a Far-Infrared 
Survey with \emph{Herschel} (KINGFISH; Kennicutt et al.\ 2011) program and
are described in detail in Engelbracht et al.\ (2010), Sandstrom et al.\ 
(2010), and Dale et al.\ (2012). Both Photodetector Array Camera and 
Spectrometer (PACS; Poglitsch et al.\ 2010) and
Spectral and Photometric Imaging REceiver (SPIRE; Griffin et al.\ 2010)
observations of NGC\,1291 were acquired.  PACS imaging was obtained in
scan mode at the medium scan speed of $20\arcsec$\,s$^{-1}$.  The
integrations achieved per pixel lead to approximate $1\sigma$ surface
brightness sensitivities of $\sigma_{\rm sky} \sim 7$, 7 and 3\,MJy sr$^{-1}$
at 70, 100, and 160\,$\micron$, respectively.  
SPIRE data were taken in Large-Map mode to 1.5 times the optical radius to 
depths of 3.2, 2.5, and 2.9\,mJy\,beam$^{-1}$,
where the beam sizes are 423, 751, and 1587 square arcseconds, at 250, 350, 
and 500\,$\micron$, respectively.
PACS and SPIRE images are the product of reduction with {\it Herschel}
Interactive Processing Environment (HIPE) version 5.0.0 (Ott 2010).
PACS data were also analyzed with Scanamorphos (Roussel et al.\ 2010; see
Section 3).  For PACS and SPIRE data,
there is an additional modified background subtraction performed, in 
which the areas 
containing bright objects in the field are ignored when calculating the 
offset, but all other reductions are standard.  This masked region was 
$7\farcm8$ in radius for NGC\,1291.  NGC\,1291 is located in an 
area of the sky with low infrared cirrus emission, so further image 
manipulation is not necessary.  
Figure 1 shows the final PACS, MIPS, and SPIRE images of NGC\,1291.

\begin{figure}
\plotone{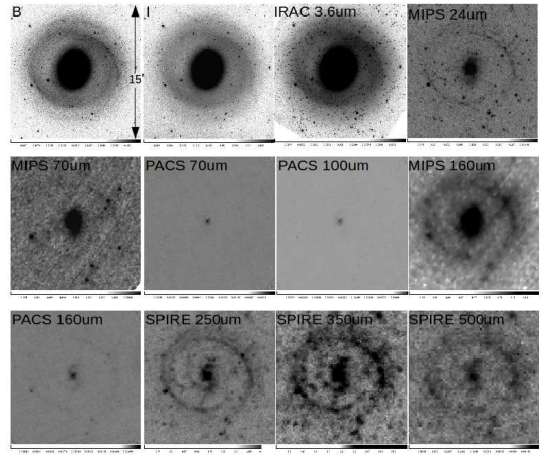}
\caption{Images of NGC\,1291.  
North is up and east is to the left.  Starting
in the top left corner the wavelengths are:  $B$, $I$, IRAC 3.6\,$\micron$, 
MIPS 24\,$\micron$, MIPS 
70\,$\micron$, PACS 70\,$\micron$, PACS 100\,$\micron$, MIPS 160\,$\micron$,
PACS 160\,$\micron$, SPIRE 250, 350 and 500\,$\micron$.}
\end{figure}

{\it Spitzer} photometry used in this work is from the SINGS project 
compiled by Dale et al.\ (2007).  PACS and 
SPIRE photometry is performed with a simple circular aperture, chosen to
encompass the galaxy's mid-infrared light, after
a median background subtraction determined from blank sky pixels away from the
galaxy.  All PACS and SPIRE photometric
points are given a 15\% uncertainty, including uncertainty due to the
background subtraction, a 1\% beam size uncertainty and calibration 
uncertainties (5\% for PACS and 7\% for SPIRE as given by the
Observer's Manuals of those instruments).

For the purposes of further analysis, we define the central or inner portion 
of the galaxy to be contained within a circle of radius $3\arcmin$ 
($\sim$9\,kpc).  We define the ring to be in the aperture between circles of 
radii of $3\arcmin$ and $6\farcm5$ ($\sim$20\,kpc) from the center.  Outside a 
radius of $6\farcm5$ is considered to be off-galaxy (sky).

\section{RESULTS}

\subsection{General Image Description}

Comparing the morphology of NGC\,1291 over a range of wavelengths allows
us to determine the spatial distribution of the various stellar populations 
with respect to the spatial distribution of the dust (see Buta et al.\ 2011
for a comprehensive review on the uses of morphology). 
Figure 1 shows that the typically identified features of the galaxy in the 
optical
are also seen at infrared wavelengths.  There is a bright inner elongated, 
oval-shaped bulge, 
an area outside this bulge with very little flux, and an outer discontinous
ring marked with dense knots of emission of varying brightnesses, which are
likely regions of recent star formation.  The
inner lens prominent at $B$-band disappears in the infrared.
In particular, the outer ring is faint relative to the bulge at bandpasses 
below 100\,$\micron$ and becomes more prominent at the 160\,$\micron$ and SPIRE
wavelengths, consistent with findings in the proceeding by Bendo 
(2006). There are localized bright spots in the ring in the northwest and
southeast arcs.

The resolution gains made by {\it Herschel} over {\it Spitzer} 
($5\farcs2$ for PACS versus $18\arcsec$ for MIPS at 70\,$\micron$) are 
illustrated by the level of detail in the central portion of the
galaxy seen in the PACS image.  Figure 2 shows the inner 
$3\arcmin\times3\arcmin$ of both images.  The PACS image has a bright
central region surrounded by diffuse emission in the form of two extensions
with a brighter patch on the northwest side.  This is not visible
in the 70\,$\micron$ MIPS image, where the lower resolution smears these 
details.  

\begin{figure}
\plotone{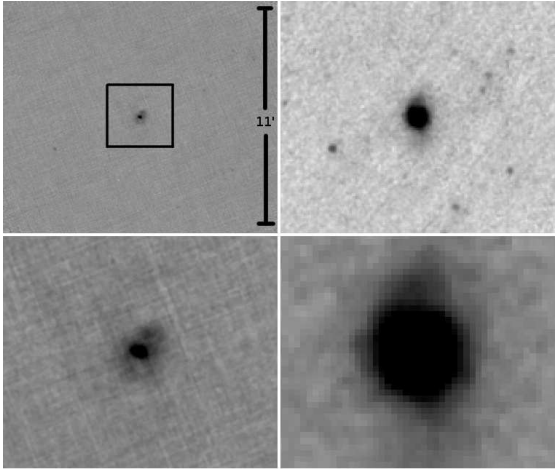}
\caption{{\it Top}:  PACS ({\it left}) and MIPS ({\it right}) 
70\,$\micron$ images of NGC\,1291.  North is up and east is to the left, and 
the field of view is 
$11\arcmin$.  The MIPS image shows the outer ring while the PACS image
does not.  The box represents the field of 
view of the corresponding bottom image.  {\it Bottom}:  The same images 
zoomed in to a $3\arcmin$ field of view, showing the resolution difference
between the two instruments.}
\end{figure}

Given the larger beam size of the MIPS instrument and the difference in the
operating temperatures of the two telescopes, it was anticipated that the 
70\,$\micron$ and 160\,$\micron$ MIPS images would be more sensitive to 
diffuse emission, such as in the outer disks of galaxies, by a factor of five 
for dust emission at T$=15$\,K and by a factor of seven for dust emission at 
T$=20$\,K for a similar number of passes by each instrument (e.g., Hinz et 
al.\ 2008a).  Figure 2 also shows a $11\arcmin\times11\arcmin$ view,
where the MIPS image has a faint detection of the ring at the northwestern
and southeastern hotspots that PACS does not.  This may be due to the 
intrinsic instrument differences as
suggested above or due to the methods with which the PACS data 
are reduced currently.  Highpass filtering used during the production of the 
PACS maps by HIPE may remove such diffuse features.  Using the Scanamorphos 
software (Roussel et al.\ 2010; Roussel 2012) after an initial 
process with HIPE recovers some diffuse features for KINGFISH galaxies.  
This software produces maps
after subtracting the thermal drift and the low-frequency noise of each
bolometer purely by using the available redundancy.  
However, the outer ring of NGC\,1291 does not appear at 70\,$\micron$ even
when Scanamorphos is used.
In this case, we find that is helpful to use both the PACS and MIPS
complementary data sets for analysis.  A discussion of apparent discrepancies
between MIPS and PACS photometry can be found in Aniano et al.\ (2012).

\subsection{Star Formation in the Ring}

The contribution from the young stellar population to the emission from the
ring can best be
seen through a combination of ultraviolet imaging, with the NUV thought to 
trace the
few hundred Myr stellar population, H\,$\alpha$ imaging, tracing the 
$\le10$\,Myr population, and infrared images at 24\,$\micron$,
representing the warm dust emission generated near star-forming regions.
Figure 3 shows {\it GALEX} NUV and FUV images (Gil de Paz et al.\ 2007), an
H\,$\alpha$ continuum-subtracted image (SINGS Legacy ancillary data) and the
MIPS 24\,$\micron$ image (SINGS; Dale et al.\ 2007). 
NGC\,1291 is not classified as an extended ultraviolet disk by Thilker et 
al. (2007) since the UV morphology is coincident with the optical morphology 
of the ring; this coincidence implies that there is a more evolved stellar 
component in the ring in addition to the young stars.
The ring in H\,$\alpha$ and 24\,$\micron$ is very thin ($\sim40\arcsec$ across
at 24\,$\micron$) and is discontinuous, 
with discrete sources appearing mostly in arcs on the northwest and
southeast sides of the galaxy.  The UV images show many more star formation
knots to the north and west of the main ring than the other wavelengths.

\begin{figure}
\plotone{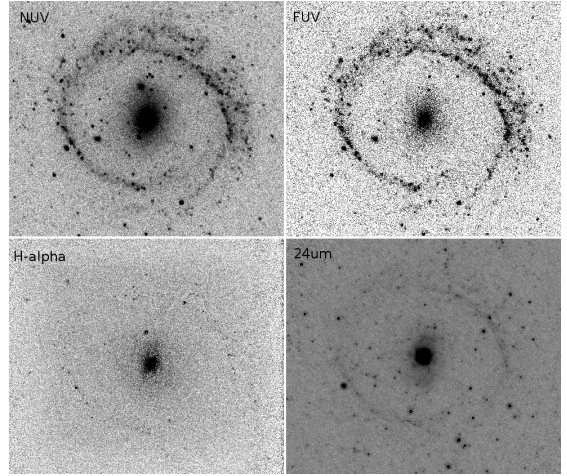}
\caption{Images of NGC\,1291 in the NUV and FUV, continuum-subtracted H\,$\alpha$, and 24\,$\micron$.  North is up and east is to the left.}
\end{figure}

In a detailed study of H\,{\sc ii} regions in 
early-type ringed galaxies by Crocker et al.\ (1996), NGC\,1291 is 
considered to be the most unusual galaxy of their sample in H\,$\alpha$ 
emission.  They describe a diffuse
and spiral-like H\,$\alpha$ distribution in the inner region of the galaxy
which appears uncorrelated with the bar and lens features visible in optical 
images, with differing shape and orientation.  The central emission
also takes on this spiral-like appearance at 24\,$\micron$.
Additionally, the H\,$\alpha$ emission near the galaxy center is 
described by Crocker et al.\ (1996) as asymmetric, and weak H\,$\alpha$ arms
extend beyond the inner lens.  These regions are filled with diffuse 
patterns of ionized gas filaments to a radius of $\sim5$\,kpc, which is 
analogous to a pattern seen in the later type galaxy M\,31 (Ciardullo et al.\
1988) and M\,81 (Devereux et al.\ 1995), and assumed to be caused by the 
ejection of the ISM via a galactic wind.
The diffuse and unusual H\,$\alpha$ emission and the H\,{\sc i} observations, 
showing an outer ring with a pronounced central hole (van Driel et al.\ 1988), 
have led to supposition that
there is little to no star formation in the central bulge (Hogg et al.\ 2001).
However, we now see that the
inner portion of the galaxy appears to be brighter compared to the ring at
both H\,$\alpha$ and 24\,$\micron$, making it unlikely that there is no star
formation occurring centrally.  

Global star formation rate estimates for NGC\,1291 differ slightly with 
method.  
The rates are log\,SFR(M$_{\odot}$\,yr$^{-1}$) $ = -0.5\pm0.4$ (H\,$\alpha$; Kennicutt
et al.\ 2008), $-0.45\pm0.04$ (FUV+TIR; Skibba et al. 2011), 
and $-0.07\pm0.05$ 
(H\,$\alpha$+24\,$\micron$; Calzetti
et al.\ 2010).  While these star formation rates are within the known range
for such galaxies  (Kannappan et al.\ 2009; Skibba et al.\ 2011; 
Crocker et al.\ in prep.), the H\,{\sc i} gas mass of NGC\,1291 is higher than
expected from its morphological type and is not coincident with any central
star formation.

To understand the star formation rates specifically in the ring, 
we calculate rates using the {\it GALEX} UV data for this portion of the 
galaxy only.  The FUV and NUV total magnitudes, within a circle of radius 
$6\farcm5$, are found
to be 13.5 and 14.4, respectively, in agreement with the {\it GALEX} 
Atlas of Nearby Galaxies
(Gil de Paz et al.\ 2007).  For the aperture between $3\arcmin$ and $6\farcm5$
that contains only the ring, the NUV magnitude is 15.1, which translates to a
log star formation rate (M$_{\odot}$\,yr$^{-1}$) = -0.52, following Donovan
et al. (2009), who use a relation from Kennicutt (1998) for a similar 
early-type galaxy with an ultraviolet ring.  Given the {\it GALEX} resolution
and the distance to the galaxy and following Kennicutt (1989), we then 
calculate an average star formation rate density
for the ring of $6.9\times10^{-5}$\,M$_{\odot}$\,yr$^{-1}$\,kpc$^{-2}$, with 
no corrections for reddening or dust extinction.  Foreground extinction is 
estimated to be small
here, only $A_U = 0.07$\,mag at $U$-band (Schlegel et al.\ 1998).  
Even if the extinction is
2-3 times larger than this estimate in the UV, the revised log star formation 
rate (M$_{\odot}$\,yr$^{-1}$) is only 
$-1.44$, for a star formation rate density of 
$8.3\times10^{-5}$\,M$_{\odot}$\,yr$^{-1}$\,kpc$^{-2}$.  This density is quite
low for an outer ring (though similar to those derived for ESO\,381-47 and 
NGC\,404; 
Donovan et al.\ 2009; Thilker et al.\ 2010) and may be due to the fact that 
the UV emission is discontinuous across the large area
of the ring (i.e., between $3\arcmin$ and $6\farcm5$).


\subsection{Cool Dust in the Ring}

Summing the sky-subtracted fluxes in the inner galaxy and in the ring area in
each band and taking their ratio yields a simple measurement of the change in 
brightness of the central portion of the galaxy relative to the brightness of 
the ring.  Figure 4 shows these ratios from 24 to 500\,$\micron$, where the
errors are calculated from photometric uncertainties.  We see that
the ratio of the flux in the ring aperture to 
the flux in the central circle increases with increasing wavelength,
indicating that there is a corresponding decrease in temperature of the dust
in the ring compared to the inner galaxy.

\begin{figure}
\plotone{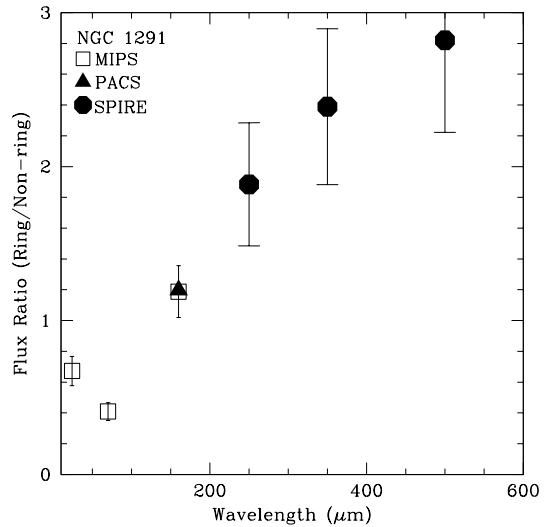}
\caption{The ratio of the flux contained in the ring to the 
flux contained
in the inner portion of the galaxy versus wavelength.  The central portion
of the galaxy is that within a $3\arcmin$ radius of the center.  The
ring is taken as the area between $3\arcmin$ and 6$\farcm$5 from the center.
Points are represented from PACS, MIPS and SPIRE.  Error bars are based on
estimated photometric errors as stated in the text.}
\end{figure}

Similarly, Figure 5 shows
the SED of the whole galaxy at far-infrared wavelengths, along with the
SED points for the inner portion of the galaxy only (within $3\arcmin$) and for
the ring only (between $3\arcmin$ and $6\farcm5$).  The SED for the ring
shows that this area generates cooler emission than the central region of
the galaxy.  The central portion of the galaxy contributes more to the total
galactic emission from warm dust at 24\,$\micron$ ($\sim60\%$) but
is quickly overtaken by the cool ring starting at 160\,$\micron$, as already
suggested by the results of Figure 4.
To calculate the temperatures of the emitting dust in various portions of
the galaxy,
we fit a blackbody with a frequency-dependent emissivity with an exponent of 
1.5 to both the
ring and inner galaxy SEDs from 70 to 500\,$\micron$, 
following Engelbracht et al.\ (2010) and Skibba et al.\ (2011).  
Trends are unaffected if emissivity values
of 1 or 2 are used, although the temperatures systematically increase or 
decrease, respectively.  The
uncertainties in the temperatures are computed via a Monte Carlo simulation
in which 10,000 trials are performed.  The photometric measurements are
allowed to vary in a normal distribution with a standard deviation indicated
by the photometric uncertainty.
The cool dust temperature of the ring
is $T_{\rm ring}=19.5\pm0.3$\,K and for the inner portion of the galaxy it is
$T_{\rm inner}=25.7\pm0.7$\,K.  This gives $T_{\rm inner}/T_{\rm ring}\sim1.3$.

\begin{figure}
\plotone{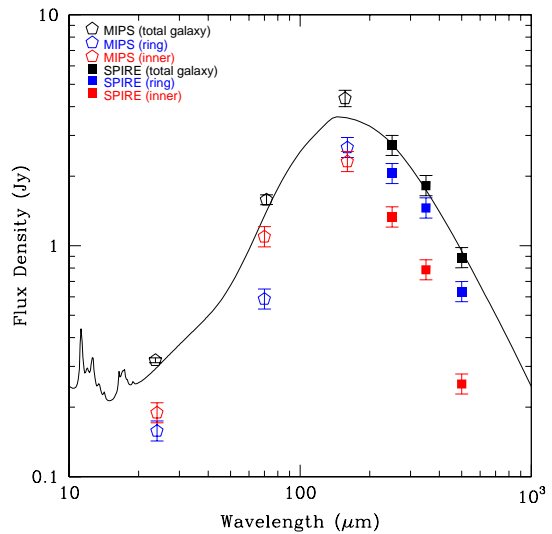}
\caption{SEDs of the inner portion of NGC\,1291, the outer ring of NGC\,1291,
and NGC\,1291 in total.  Open pentagons in
all colors are MIPS data points.  Solid squares in all colors are SPIRE data
points.  Black points represent the total SED of the galaxy.  Red points 
represent the SED of the
inner portion of the galaxy, while blue points represent the SED of the ring
only.  The black solid line is a model fit to the total SED (all black points)
using results from Aniano et al.\ (2012).}
\end{figure}

Both of these dust temperature
values are typical of cool dust temperatures in nearby galaxies (e.g., Hinz
et al.\ 2006; 2008b).  The values
for NGC\,1291 are slightly different from those found by 
Engelbracht et al.\ (2010), likely due to the difference in apertures used,
but consistent with their overall finding that earlier type spirals have
higher central temperatures by 20\% to 50\% over their late-type counterparts.
The temperature ratio is closer to that
found for strongly barred galaxies compared to those derived in
weakly barred galaxies, though the
bar classification for this galaxy changes slightly between optical and 
near-infrared analyses, from SB to SAB (Buta et al.\ 2010).

The dust masses for NGC\,1291, as a whole and for its components, can be 
calculated using the mid-infrared photometry from IRAC, 
PACS, MIPS and SPIRE and the 
models of Draine \& Li (2001; 2007) and constructed as suggested in Draine et 
al.\ (2007), using a mixture of amorphous silicate grains and carbonaceous 
grains with a distribution of grain sizes.  All images are convolved 
to the MIPS 160\,$\micron$ resolution before modeling (Aniano et al.\ 2011).  
Details of the use of these models 
specifically with {\it Herschel} photometry are described extensively in Aniano 
et al.\ (2012) for two other spiral galaxies from KINGFISH. 
The results of such fitting for NGC\,1291 are summarized here, with the 
black solid line in Figure 5 representing the final fit to the data for the
whole galaxy.  The apertures, dust masses ($M_{\rm dust}$), aperture areas, 
mean starlight heating intensities ($<U>$), fluxes radiated by dust 
corresponding to photodissociation 
regions ($f_{PDR}$) and the fractions of grain mass contributed by polycyclic 
aromatic hydrocarbons ($q_{PAH}$) are listed in Table 1.
The total dust mass for the galaxy, calculated from a sum of the modeling for
each individual galaxy pixel, is $M_{\rm dust}=2.8\times10^7\pm3.5\times10^6$\,M$_{\odot}$,
with a total dust luminosity of $L_{\rm dust}=2.4\times10^9\pm2.3\times10^7$
\,L$_{\odot}$.
Dust masses calculated for annular apertures with radii in intervals
of $1\arcmin$ out to $6\arcmin$ show that the dust in the ring dominates the
total dust mass of NGC\,1291.  The average starlight heating intensity is 
calculated
to be the largest in the inner portion of the galaxy.  This $<U>$ can be
related to the characteristic dust temperature, $T_{\rm dust}$, by the 
approximate relation $T_{\rm dust} = 18{\rm K} \times <U>^{1/6}$, where
$<U>$ is in units of the starlight energy density in the solar
neighborhood (Aniano et al.\ 2012).

\section{DISCUSSION}
Several explanations for the existence of an outer ring with cooler dust may
be plausible.  It is possible that all spiral galaxies have a dust
temperature gradient between their inner and outer disks.  This might be due 
to the concentration of old stars or star formation
occurring at the center of the galaxy in comparison to the more sparse 
stellar population of the comparatively large outer disk, i.e., that the 
density of
the radiation field in the center of a galaxy is higher than that in the outer
parts.  Another explanation could be that the nucleus
of the galaxy contains an AGN which heats the dust to higher temperatures in
the inner region in comparison to the outskirts.  Perhaps the inner bar of the
galaxy fuels star formation in that area by funneling gas continuously, such
that rapid star formation heats the inner dust to warm temperatures.
Or it could be that the stellar 
population of the outer ring is older than that of the inner features. In
this case, the older stars maintain the dust at cooler temperatures than areas 
with active star formation.  We now discuss each of these possible 
explanations in turn.

\subsection{Temperature Gradients}

First, we explore the idea that all galaxies have similar dust temperature
gradients to NGC\,1291.  Engelbracht et al.\ (2010) showed that 
the central areas of early-type spiral galaxies generally have enhanced dust 
heating of their cool dust components compared to their disks.  They 
find that, on average, the cool dust temperature of the central component is 
$15\pm3\%$ hotter than the disk.  Similarly, Pohlen et al.\ (2010) showed
that SPIRE surface brightness ratios seem to decrease with radius in spirals, 
implying that the dust in the outer regions is colder than dust in the centers
of galaxies.  Galametz et al.\ (2012; in prep.) study 11 galaxies in the
KINGFISH sample, including NGC\,1291, fitting {\it Spitzer} and {\it Herschel}
data SEDs with two modified blackbodies and creating spatially-resolved maps 
of their
dust properties.  They also see systematic drops in dust temperature
with radius for disk galaxies, also on the order of 10-15\,K from inner to outer
galaxy.

However, it is not obvious that NGC\,1291 should follow this pattern and that
outer rings should be dominated by cool dust emission.  
UV rings in S0 and early-type spiral galaxies are often 
associated with sites of recent star formation, sometimes marked by
clumpy H\,{\sc ii} regions, and can even be the only sites of star formation
in some galaxies (e.g., Donovan et al.\ 2009; Thilker et al.\ 2010).  A recent 
study of five nearby barred S0 galaxies with rings by Marino et al.\ (2011) 
showed that these outer 
rings account for a majority (up to 70\%) of the flux at UV
wavelengths, indicating their young stellar population.  
Buta et al.\ (2010) specifically discuss the morphology of NGC\,1291
based on $B$-band and 3.6\,$\micron$ IRAC images (Sheth et al.\ 2010), saying 
that the outer ring is ``where most of the
recent star formation is taking place'' and is prominent in the $B$-band 
(Figure 1).
At 3.6\,$\micron$, however, the ring does not stand out and is instead a 
broad ellipse in a rounder diffuse background (Figure 1 and their Figure 7).
Therefore, we might expect that, if the 24\,$\micron$ emission is associated 
with the warm (T$\sim50$\,K) dust heated by the young stellar
population (seen at UV and $B$-band to be prominent), then the
warm dust at 24\,$\micron$ should be brighter in the ring as well, yet 
the wavelengths associated with cool (T$\sim20$\,K) dust are brighter relative 
to the flux from the rest of the galaxy.

To study the dust temperature gradient in NGC\,1291 in more detail, Figure 6 
displays circles of radii $2\arcmin$, $3\arcmin$, $4\arcmin$, $5\arcmin$, and 
$6\arcmin$ overlaid on the SPIRE 250\,$\micron$ image, along with the 
estimated temperature of the dust within each annular aperture, using the same 
blackbody fitting technique as before.  (These annuli match the dust mass
annuli calculations found in Table 1.  Unlike Table 1, the data for Figure
6 are not convolved to the 160\,$\micron$ resolution.)  We remind the reader
that such temperatures are representative, characterizing large grains that
dominate the emission at wavelengths greater than 70\,$\micron$, and that, in
fact, such large annuli contain a wide range of temperatures, including small
grains undergoing temperature fluctuations.  Keeping this in mind, 
if all disks, with or without rings, have similar dust temperature gradients
due to the gradually increasing distance between star formation regions with
radius, then we should be able to identify the same drop in temperature for 
spirals other than NGC\,1291.  We choose, from the KINGFISH sample, NGC\,628 
(SAc; $d=7.3$\,Mpc), which does not contain any rings but is face-on,
is at a similar distance to NGC\,1291 and has a similar angular size
and total infrared
luminosity to NGC\,1291.  We show in Figure 6 the dust
temperatures for this galaxy for the same annular apertures as NGC\,1291.
For this galaxy, we see a gradual
decline in temperature with some scatter.  We see neither the dramatic dip
in temperature at $3\arcmin$ that is seen for NGC\,1291, nor do we see such a
large spread in temperatures, with NGC\,628 having a temperature difference
between inner and outer areas of 
4.8\,K versus 8.2\,K for NGC\,1291. Thus, it appears that the dust temperature
from the ring is cooler than would be expected from a simple 
temperature decline with radius for a normal (non-ringed) spiral galaxy.  
Comparisons of spatially resolved dust temperatures with a larger number of 
KINGFISH galaxies are shown in Galametz et al.\ (2012).

\begin{figure}
\plottwo{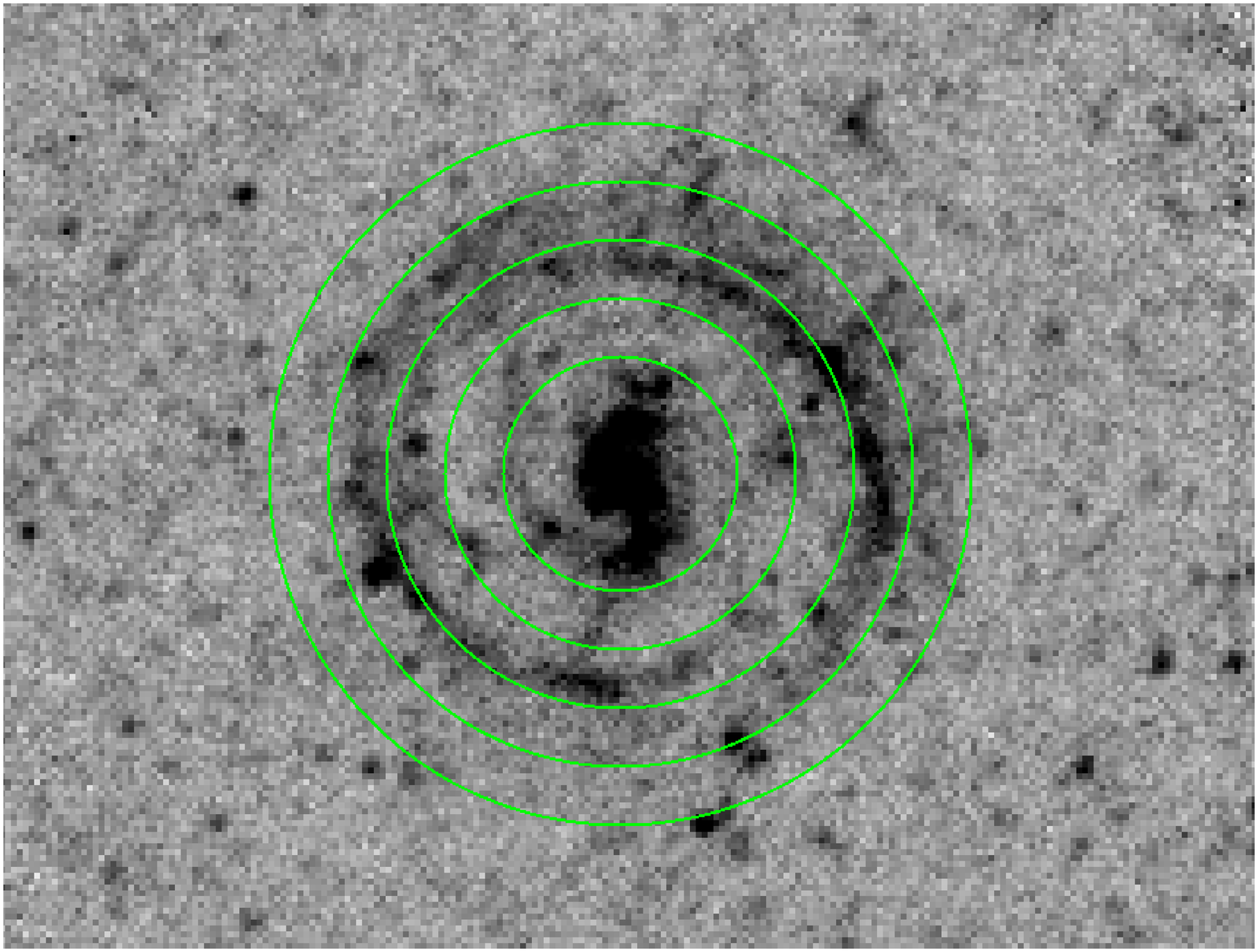}{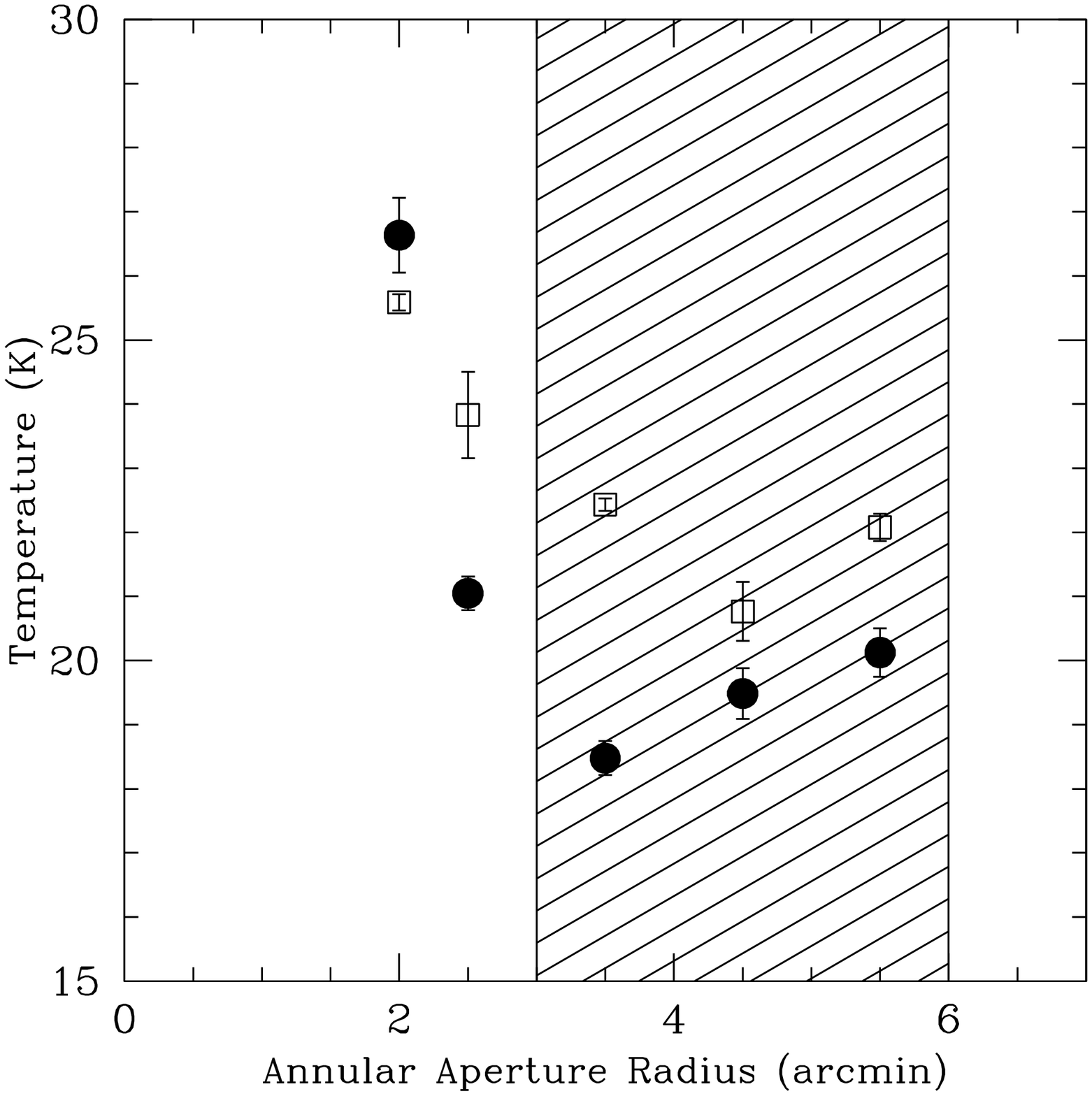}
\caption{{\it Left:} The 250\,$\micron$ image of NGC\,1291 (grayscale)
overlaid with 
apertures at $2\arcmin$, $3\arcmin$, $4\arcmin$, $5\arcmin$ and $6\arcmin$  
radii (green circles).
{\it Right:} The dust temperature deduced from the infrared flux densities 
within
each aperture.  
Solid dots represent NGC\,1291 while open squares represent
NGC\,628, a comparison face-on spiral galaxy.
The shaded region indicates the approximate area occupied
by the ring of NGC\,1291 at SPIRE wavelengths.}
\end{figure}

\subsection{AGN Heating}

The next possibility is that non-stellar sources such as an AGN could heat
the dust in the inner portion of the galaxy, where we find a warmer dust
temperature compared to the outer ring.  This is suggested by the fact that
outer rings are found to be much more frequent in galaxies
with Seyfert nuclei (Hunt \& Malkan 1999).  NGC\,1291 is an X-ray source,
with emission detected within $2\farcm5$ (Bregman et al.\ 1995), which appears
to be consistent with two energy components with their origin in a stellar
component and an extended hot gas component (Hogg et al.\ 2001). Irwin et
al.\ (2002) found an excess of soft emission similar to that seen in several 
other low-luminosity AGNs.
Moustakas et al.\ (2010) classify NGC\,1291 as an ``AGN'' type galaxy
based on optical spectra of the nuclear region, although they note that one
or more emission line(s) failed their S/N$>2$ requirement for such a 
classification.  Recent analysis of X-ray spectra of the nuclear source
also indicates a low-luminosity AGN with moderate obscuration (Luo et 
al.\ 2012) which could be a source of dust heating.


\subsection{Transport via the Bar}

The third possibility is that the bar in NGC\,1291, seen at the various 
wavelengths in Figure 1, funnels gas into the inner region of the galaxy,
replenishing the supply necessary for star formation.  This compact region of
star formation could then generate warmer dust temperatures than in the outer
ring (see also Engelbracht et al.\ 2010).  
If gas is being transported inwards in this manner, the observational
evidence would likely be in the form of ``hotspots'' at the ends of the bar -
features that are brighter than the rest of the bar formed by shock heating
of the gas as it is drawn to the center of the galaxy, causing a ``pile-up''
near the inner Lindblad resonance (Combes \& Gerin 1985; 
Shlosman et al.\ 1989).  This is not seen in the
H\,{\sc i} data, which has a hole at the center (van Driel et al.\ 1988), or
in the other wavelengths that are available.  However, we cannot rule out
this possibility, as such hotspots are thought to be small, very faint 
features. Hunt \& Malkan (1999) argue that rings alone could be signs of inward
transport of material, essentially a second ``pile-up'' of dust and gas in the 
outer resonance of the galaxy; the star formation in the ring prominent in the 
UV, presumably indicating stellar ages of $\sim100$\,Myr, could be a residual
signature of such a build-up. We note that transport of material via the
bar may be related to the AGN hypothesis in the previous section, in that 
AGNs as well as star formation can be fueled via inflow.

\subsection{Stellar Age Gradient}

The last explanation for the change in cool dust temperature between the
inner galaxy and outer ring is that the ring is composed mainly of an old
population of stars with some current star formation.
In this case, the dust in the ring was produced by generations of
stars in the past and is heated now mainly by the old ($\sim10$\,Gyr)
stellar population, 
shown best by the 3.6\,$\micron$ image (Buta et al.\ 2010 and Figure 1) where 
the ring is a large, diffuse entity spanning much more physical
space than seen at far-infrared wavelengths.
Such ``old population rings'' have been suggested by comparisons
between $B$ and $I$-band images of nearby galaxies.  Buta (1995) showed a 
typical galaxy of this type (IC\,1438) which has a nuclear ring, an oval 
inner 
ring, and an outer ring that is much brighter in the $I$ passband than in the 
$B$.  The explanation given for this difference in brightness was that the 
outer ring component formed first and left
behind a stellar remnant as the other two inner features formed.  
An old outer ring would be unusual in that the dynamical timescales for
forming rings or arms are much shorter at smaller disk radii (e.g., Freeman
et al.\ 2010).

A remnant of
this type could be the source of dust heating for the emission we see at 
far-infrared wavelengths.  In addition, there would also be a contribution to 
the dust heating from ionizing ultraviolet photons which have escaped from 
the existing (but weaker, in terms of contribution) star forming regions.
This combination of dust heating sources has been well modeled by, e.g.,
Misiriotis et al.\ (2001) and Popescu et al.\ (2002; 2011), and proposed to 
explain observations of other nearby galaxies 
(Hinz et al.\ 2004, Tabatabaei et al.\ 2007, Bendo et al.\ 2010).  (Though see
Groves et al. 2012 for a case where the old stellar population heats dust to
warm temperatures.)
Inspection of the $B$-$I$ image (R.\ Buta, private communication) shows such
a color difference and will be investigated further in Buta et al.\ (2012).
Simulations by Freeman et al.\ (2010) addressing Hoag's 
object, an outer ring galaxy with no inner morphological features, show that 
an outer ring can be formed via perturbations from a bar which then
dissipates slowly over time. In the environment of these instabilities,
the majority of the simulated gas particles fall to the center, while the 
remaining particles create an outer ring, with the region between the two
essentially empty. If this is the manner in which NGC\,1291 is forming its
ring, then the implication is that the ring is unlikely to be older than
the inner regions.

Noll et al.\ (2009) use SED fitting techniques to study the star-formation
history of NGC\,1291 and posit two exponential starburst events for the 
galaxy, one of which occurred 10\,Gyr ago and the second of which occurred 
200\,Myr ago.  This analysis was performed for the entire galaxy.  A similar
study for the inner portion and the ring separately would help disentangle
their stellar histories.

\section{SUMMARY}

Far-infrared images from {\it Herschel} of the ringed galaxy NGC\,1291 show
that the morphology of the galaxy changes from that found in the optical,
with some features such as the lens disappearing, and other features such
as the spiral-like inner region identified for the first time.
Additionally, the large outer ring becomes more prominent at 
wavelengths longer than 160\,$\micron$.  An 
exploration of temperature of the dust emission via blackbody fitting in 
annular apertures over the face of the galaxy yields the result that the dust 
in the ring is $\sim6$\,K cooler than that found in the inner $\sim3\arcmin$.  
Further modeling shows that this cool dust in the ring dominates the total 
galaxy dust mass.  We discuss whether, despite the 
presence of recent star formation traced by UV emission, the dust in the ring 
is heated primarily
by a large, diffuse stellar population which is older than that of the inner 
portion of the galaxy and the bulge, lens and bar features located there.  
This older stellar population would allow the cool dust to remain at 
temperatures of $\sim20$\,K.  Other possibilities for the dust temperature
difference discussed here include an expected temperature gradient due to
the change in starlight heating intensity, heating of the inner region by
a low-luminosity AGN, and increased star formation in the inner region due
to gas transport along the bar.

\acknowledgments

J.~L.~H.\ thanks R.\ Buta for a preview of his $B$ and $I$ images of
NGC\,1291 and for helpful discussions.
The following institutes have provided hardware and software elements to the SPIRE project: University of Lethbridge, Canada; NAOC, Beijing, China; CEA Saclay, CEA Grenoble, and OAMP in France; IFSI, Rome, and University of Padua, Italy; IAC, Tenerife, Spain; Stockholm Observatory, Sweden; Cardiff University, Imperial College London, UCL-MSSL, STFC-RAL, UK ATC Edinburgh, and the University of Sussex in the UK. Funding for SPIRE has been provided by the national agencies of the participating countries and by internal institute funding: CSA in Canada; NAOC in China; CNES, CNRS, and CEA in France; ASI in Italy; MCINN in Spain; Stockholm Observatory in Sweden; STFC in the UK; and NASA in the USA. Additional funding support for some instrument activities has been provided by ESA.
This research has made use of the NASA/IPAC Extragalactic Database (NED) which is operated by the Jet Propulsion Laboratory, California Institute of Technology, under contract with the National Aeronautics and Space Administration.

\clearpage

\begin{deluxetable}{clllllllllllll}
\tablecaption{Dust Masses and Dust Mass Weighted Starlight Heating Intensities}
\tablewidth{350pt}
\tablehead{
\colhead{Annular} & \colhead{M$_{dust}$} &  \colhead{Area} & \colhead{$<U>$} & \colhead{f$_{PDR}$} & \colhead{q$_{PAH}$} & \\ \colhead{Aperture}
\linebreak & & & & & &\\ \colhead{(Arcmin)}
\linebreak & \colhead{(M$_{\odot}$ kpc$^{-2}$)} & \colhead{(kpc$^2$)} & \colhead{(U$_{\odot}$)} & \colhead{(\%)} & \colhead{(\%)}}
\startdata
0-1 & 3.1E$+04\pm1.7$E+03 & 28.8  & $4.6\pm0.3$ & $9.38\pm1.14$ & $1.60\pm0.20$ &\\
1-2 & 2.1E$+04\pm1.1$E+04 & 86.3  & $1.7\pm0.6$ & $7.72\pm2.05$ & $1.30\pm0.80$ &\\
2-3 & 1.8E$+04\pm7.9$E+02 & 143.8 & $0.7\pm0.2$ & $8.69\pm5.14$ & $1.30\pm1.20$ &\\
3-4 & 3.1E$+04\pm5.7$E+03 & 201.3 & $0.5\pm0.1$ & $7.76\pm1.13$ & $2.40\pm0.70$ &\\
4-5 & 4.1E$+04\pm1.4$E+04 & 258.8 & $0.4\pm0.1$ & $7.49\pm3.59$ & $3.00\pm1.30$ &\\
5-6 & 2.0E$+04\pm7.9$E+03 & 316.3 & $0.4\pm0.1$ & $9.15\pm1.54$ & $2.10\pm0.20$ &\\
\enddata
\end{deluxetable}


\begin{references}

\reference{}Aniano, G., Draine, B.~T., Gordon, K.~D., \& Sandstrom, K.\ 2011, \pasp, 123, 1218 
\reference{}Aniano, G., et al.\ 2012, accepted to ApJ
\reference{}Appleton, P. N. \& Struck-Marcell, C.\ 1996, Fundam.\ Cosm.\ Phys., 16, 111
\reference{}Bagley, M., Minchev, I., \& Quillen, A.~C.\ 2009, \mnras, 395, 537 
\reference{}Bendo, G.~J.\ 2006, Astronomical Society of the Pacific Conference Series, 357, 192 
\reference{}Bendo, G.~J., et al.\ 2010, \aap, 518, L65 
\reference{}Bregman, J.~N., Hogg, D.~E., \& Roberts, M.~S.\ 1995, \apj, 441, 561 
\reference{}Brook, C., Richard, S., Kawata, D., Martel, H., \& Gibson, B. K.\ 2007, \apj, 658, 60 
\reference{}Buta, R.\ 1995, \apjs, 96, 39
\reference{}Buta, R., \& Combes, F.\ 1996, \fcp, 17, 95
\reference{}Buta, R.~J., et al.\ 2010, \apjs, 190, 147 
\reference{}Buta, R.~J.\ 2011, arXiv:1102.0550
\reference{}Buta, R.~J.\ 2012, in preparation
\reference{}Caldwell, N., Kennicutt, R., Phillips, A.~C., \& Schommer, R.~A.\ 1991, \apj, 370, 526 
\reference{}Calzetti, D., et al.\ 2010, \apj, 714, 1256 
\reference{}Ciardullo, R., Rubin, V.~C., Jacoby, G.~H., Ford, H.~C., \& Ford, W.~K.\ 1988, \aj, 95, 438
\reference{}Combes, F., \& Gerin, M.\ 1985, \aap, 150, 327 
\reference{}Crocker, D.~A., Baugus, P.~D., \& Buta, R.\ 1996, \apjs, 105, 353 
\reference{}Dale, D.~A., Aniano, G., Engelbracht, C.~W., et al.\ 2012, \apj, 745, 95 
\reference{}Dale, D.~A., et al.\ 2007, \apj, 655, 863 
\reference{}de Jong, R.~S., et al.\ 2007, \apjl, 667, L49 
\reference{}de Vaucouleurs, G.\ 1975, \apjs, 29, 193 
\reference{}de Vaucouleurs, G., de Vaucouleurs, A., Corwin, H.~G., Jr., Buta, R.~J., Paturel, G., \& Fouque, P.\ 1991, Third Reference Catalogue of Bright Galaxies (New York:  Springer)
\reference{}Devereux, N., Jacoby, G.~H., \& Ciardullo, R.\ 1995, \aj, 110, 1115
\reference{}Donovan, J.~L., et al.\ 2009, \aj, 137, 5037
\reference{}Draine, B.~T., \& Li, A.\ 2001, \apj, 551, 807 
\reference{}Draine, B.~T., \& Li, A.\ 2007, \apj, 657, 810 
\reference{}Draine, B.~T., et al.\ 2007, \apj, 663, 866 
\reference{}Engelbracht, C.~W., et al.\ 2004, \apjs, 154, 248
\reference{}Engelbracht, C.~W., et al.\ 2007, \pasp, 119, 994 
\reference{}Engelbracht, C.~W., et al.\ 2010, \aap, 518, L56
\reference{}Fazio, G.~G., et al.\ 2004, \apjs, 154, 10 
\reference{}Freeman, T., Howard, S., \& Byrd, G.~G.\ 2010, Celestial Mechanics and Dynamical Astronomy, 108, 23
\reference{}Gil de Paz, A., et al.\ 2005, \apjl, 627, L29 
\reference{}Gil de Paz, A., et al.\ 2007, \apjs, 173, 185
\reference{}Gordon, K.~D., et al.\ 2007, \pasp, 119, 1019 
\reference{}Griffin, M.~J., et al.\ 2010, \aap, 518, L3 
\reference{}Groves, B., et al.\ 2012, accepted to \mnras, arXiv:1206.2925
\reference{}Hinz, J.~L., et al.\ 2004, \apjs, 154, 259
\reference{}Hinz, J.~L., et al.\ 2006, \apj, 651, 874
\reference{}Hinz, J., Gordon, K., \& Rieke, G.\ 2008a, Spitzer Proposal ID \#489, 489 
\reference{}Hinz, J.~L., Rieke, M.~J., Rieke, G.~H., et al.\ 2008b, \apj, 663, 895
\reference{}Hogg, D.~E., Roberts, M.~S., Bregman, J.~N., \& Haynes, M.~P.\ 2001, \aj, 121, 1336
\reference{}Hunt, L.~K., \& Malkan, M.~A.\ 1999, \apj, 516, 660
\reference{}Irwin, J.~A., Sarazin, C.~L., \& Bregman, J.~N.\ 2002, \apj, 570, 152 
\reference{}Kannappan, S.~J., Guie, J.~M., \& Baker, A.~J.\ 2009, \aj, 138, 579 
\reference{}Kennicutt, R.~C., Jr.\ 1989, \apj, 344, 685 
\reference{}Kennicutt, R.~C., Jr.\ 1998, \apj, 498, 541
\reference{}Kennicutt, R.~C., Jr., et al.\ 2003, \pasp, 115, 928
\reference{}Kennicutt, R.~C., Jr., Lee, J.~C., Funes, S.~J., Jos{\'e} G., Sakai, S., \& Akiyama, S.\ 2008, \apjs, 178, 247 
\reference{}Kennicutt, R.~C., Calzetti, D., Aniano, G., et al.\ 2011, \pasp, 123, 1347
\reference{}Li, J.-T., Wang, Q.~D., Li, Z., \& Chen, Y.\ 2011, \apj, 737, 41 
\reference{}Luo, B., Fabbiano, G., Fragos, T., et al.\ 2012, arXiv:1202.2358
\reference{}Marino, A., et al.\ 2011, arXiv:1105.3812
\reference{}Masters, K.~L.\ 2005, Ph.D.~Thesis
\reference{}McConnachie, A.~W., et al.\ 2009, \nat, 461, 66 
\reference{}Meurer, G., et al.\ 2006, \apjs, 165, 307
\reference{}Michard, R., \& Marchal, J.\ 1993, \aaps, 98, 29
\reference{}Misiriotis, A., Popescu, C.~C., Tuffs, R., \& Kylafis, N.~D.\ 2001, \aap, 372, 775 
\reference{}Moustakas, J., Kennicutt, R.~C., Jr., Tremonti, C.~A., Dale, D.~A., Smith, J.-D.~T., \& Calzetti, D.\ 2010, \apjs, 190, 233 
\reference{}Mu{\~n}oz-Mateos, J.~C., Gil de Paz, A., Boissier, S., Zamorano, J., Jarrett, T., Gallego, J., \& Madore, B.~F.\ 2007, \apj, 658, 1006 
\reference{}Noll, S., Burgarella, D., Giovannoli, E., et al.\ 2009, \aap, 507, 1793
\reference{}Ott, S.\ 2010, ASP Conference Series, 434, 139
\reference{}Perrine, C.~D.\ 1922, \mnras, 82, 486
\reference{}P{\'e}rez, I., \& Freeman, K.\ 2006, \aap, 454, 165
\reference{}Planck Collaboration, Ade, P.~A.~R., Aghanim, N., et al.\ 2011, arXiv:1101.2046
\reference{}Poglitsch, A., et al.\ 2010, \aap, 518, L2
\reference{}Pohlen, M., et al.\ 2010, \aap, 518, L72
\reference{}Popescu, C.~C., Tuffs, R.~J., V{\"o}lk, H.~J., Pierini, D., \& Madore, B.~F.\ 2002, \apj, 567, 221 
\reference{}Popescu, C.~C., Tuffs, R.~J., Dopita, M.~A., et al.\ 2011, \aap, 527, A109
\reference{}Prescott, M.~K.~M., et al.\ 2007, \apj, 668, 182
\reference{}Rautiainen, P. \& Salo, H.\ 2000, \aap, 362, 465
\reference{}Rieke, G.~H., et al.\ 2004, \apjs, 154, 25 
\reference{}Roussel, H., et al.\ 2010, \aap, 518, L66
\reference{}Roussel, H., 2012, arXiv:1205.2576
\reference{}Sandstrom, K., Krause, O., Linz, H., et al.\ 2010, \aap, 518, L59 
\reference{}Schlegel, D.~J., Finkbeiner, D.~P., \& Davis, M.\ 1998, \apj, 500, 525 
\reference{}Schwarz, M.~P.\ 1981, \apj, 247, 77
\reference{}Sheth, K., et al.\ 2010, \pasp, 122, 1397 
\reference{}Skibba, R.~A., Engelbracht, C.~W., Dale, D., et al.\ 2011, \apj, 738, 89
\reference{}Shlosman, I., Frank, J., \& Begelman, M.~C.\ 1989, \nat, 338, 45 
\reference{}Smith, J.~D. et al.\ 2007, \apj, 656, 770
\reference{}Stansberry, J.~A., et al.\ 2007, \pasp, 119, 1038 
\reference{}Tabatabaei, F.~S., et al.\ 2007, \aap, 466, 509
\reference{}Thilker, D.~A., et al.\ 2005, \apjl, 619, L67 
\reference{}Thilker, D.~A., et al.\ 2007, \apjs, 173, 538
\reference{}Thilker, D.~A., et al.\ 2010, \apjl, 714, L171 
\reference{}van Driel, W., Rots, A.~H., \& van Woerden, H.\ 1988, \aap, 204, 39
\reference{}Wiebe, D.~V., et al.\ 2009, \apj, 707, 1809 

\end{references}
\end{document}